\documentclass[conf]{new-aiaa}
\usepackage[utf8]{inputenc}

\usepackage{graphicx}
\usepackage{amsmath}
\usepackage[version=4]{mhchem}
\usepackage{siunitx}
\usepackage{longtable,tabularx}
\usepackage{float}
\usepackage[svgnames]{xcolor}
\usepackage{float}
\usepackage{stackengine}

\usepackage{pdfpages}
\usepackage{graphicx}
\usepackage{subfigure}
\usepackage[belowskip=-10pt,aboveskip=0pt,justification=centering]{caption}
\usepackage{overpic}
\usepackage{xfrac}

\title{Characterization of boundary layers on isothermal and adiabatic curved surfaces of a supersonic turbine cascade}

\author{Gabriel Y. R. Hamada\footnote{PhD Candidate, Faculdade de Engenharia Mecânica,  E-mail: g265310@unicamp.br}, Hugo F. S. Lui\footnote{PhD Candidate, Faculdade de Engenharia Mecânica, E-mail: hugo.slui@gmail.com.}, William R. Wolf\footnote{Associate Professor, Faculdade de Engenharia Mecânica, Departmento de Energia, E-mail: wolf@fem.unicamp.br.}}
\affil{Faculdade de Engenharia Mecânica, Universidade Estadual de Campinas, Campinas, SP 13083-860, Brasil}
\author{Carlos Junqueira-Junior\footnote{Research Engineer, Arts et Métiers Institute of 
Technology, DynFluid laboratory; E-mail: junior.junqueira@ensam.eu.}}
\affil{DynFluid, Arts et Métiers Institute of Technology, CNAM,\\ 
151 Boulevard de l'Hôpital, 75013, Paris, France}

\begin{document}

\maketitle

\begin{abstract}
The effects of adiabatic and isothermal boundary conditions are investigated on the shock-boundary layer interactions (SBLIs) in a supersonic turbine cascade. Special attention is given to the characterization of the incoming boundary layers over the convex and concave walls of the blade and their impact in the SBLIs. Large eddy simulations (LES) are performed for an inlet Mach number of $\mathbf{M_\infty = 2.0}$ and Reynolds number based on the axial chord $\mathbf{Re = 200\,000}$. For the isothermal condition, the wall to inlet temperature ratio is $\mathbf{T_w/T_{\infty}=0.75}$, representing a cooled wall. 
Different incident shock wave topologies occur on the suction and pressure sides of the airfoil. For the former, an oblique shock impinges on the boundary layer leading to a larger separation bubble. On the other hand,  a normal shock from a Mach reflection induces a small separation region near the wall for the pressure side. 
Results are presented in terms of mean velocity and temperature contours, and the incoming boundary layers are characterized by looking at the Clauser parameter, shape factor, dilatation and turbulent kinetic energy (TKE) profiles. Inspection of the shape factors show that the adiabatic wall boundary layers are more prone to separate than the isothermal ones. This is indeed observed in the airfoil suction side, but not on the pressure side, where the flow separates in the same chord position regardless of the thermal boundary condition. This is a topic for investigation in the final version of the paper. An assessment of the dilatation and TKE profiles explains the disparities of the bubble sizes on the pressure and suction sides of the airfoil.
\end{abstract}

\section{Introduction}
\label{sec:introd}

Supersonic turbines are employed in high-speed propulsion and power generation systems \citep{PANIAGUA201465, SOUSA2017247}. In such systems, the interactions between shocks and boundary layers pose a design problem. Shock-boundary layer interactions (SBLIs) originate from shock waves that develop at the stator/rotor leading edges and impinge on the boundary layers of adjacent airfoils. The shocks can induce flow separation 
by imposing severe adverse pressure gradients to the boundary layers. Moreover, they cause significant pressure and thermal fluctuations, which can damage the structural integrity of the turbine and lower the overall machine performance \citep{DELERY1985,babinsky_harvey_2011,GAITONDE2015,Klinner2019,SPOTTSWOOD2019,Sandberg2022}.

SBLIs are characterized by low-frequency dynamics which have been extensively investigated. However, the processes that drive the bubble breathing motions are still a subject of study \cite{clemens2014}. Several physical processes have been proposed to explain the low-frequency unsteadiness in SBLIs. Some investigations have found that upstream boundary layer fluctuations cause bubble pulsations \cite{beresh2002,ganapathisubramani2007effects,ganapathisubramani_clemens_dolling_2009, porter_2019,baidya_2020}, where the passage of near-wall elongated low-speed (high-speed) structures through the separated flow causes the bubble to expand (contract). Other studies \cite{wu_martin_2008,piponniau2009,estruch2018separated,hu_2021,jenquin2023investigations} associated the mass imbalance within the recirculation zone to the mechanism that drives the bubble breathing motion. These authors proposed that fluid from the separated flow can be expelled by the shear layer, while fluid from the downstream flow can be injected into the recirculation region at the reattachment position. When the mass injection is higher, the separation bubble expands. In a different fashion, other investigations have suggested that the pulsations are caused by a global instability of the  separation bubble  \cite{Touber2009,priebe2012,adler_gaitonde_2018}.

Only a few authors have investigated thermal effects  on the dynamics of SBLIs \cite{Bernardini2016,Tong2017,Volpiani2018,ZANGENEH2021}. 
Moreover, most of the studies above have been performed in situations where the incoming boundary layer develops on flat surfaces. Curved walls, on the other hand, are found in several high-speed aerodynamic devices, such as turbomachinery, transonic airfoils, and nozzles. Some authors have explored the unsteadiness of SBLIs on curved surfaces \cite{Hartmann2013,sartor2015unsteadiness,Klinner2019,Klinner2021,Lui2022}, but the majority of these studies is related to transonic flows, which differ from supersonic arrangements in several ways, including the ability of acoustic waves to propagate upstream and influence the SBLIs \cite{clemens2014}. \citet{Lui2022} performed a wall-resolved large eddy simulation (LES) of a supersonic turbine cascade at Mach 2.0. When compared to flat plates, the curved walls of the airfoil resulted in unique shock structures. Compression waves, for example, did not coalesce into a separation shock on the airfoil suction side, despite a recirculation bubble. On the other hand, a Mach reflection was formed on the pressure side, where the airfoil surface is concave. Regardless of these discrepancies, the results showed that the characteristic frequencies of the shock/bubble motions are similar to those seen in canonical flat plates. The authors also observed that near-wall elongated structures in the incoming boundary layers drive the breathing motion of the suction side separation bubble.

In this work, wall-resolved LES are performed to investigate the effects of thermal boundary conditions on the incoming boundary layers and their SBLIs. We follow up the research group work \cite{lui2022comparison,Hamada23-aiaa} and study a flow setup consisting of a supersonic turbine cascade with inlet Mach number $M_\infty= 2.0$ and Reynolds number $Re=200,000$, based on the axial chord. The thermal boundary conditions analyzed include adiabatic and isothermal walls, where for the latter the airfoil surface is cooled with $T_w/T_\infty = 0.75$. Here, $T_w$ and $T_\infty$ are the wall and inlet temperatures, respectively. The present study provides insights of the SBLIs in a stator cascade, a configuration that is more complex than the canonical compression ramp and the frequently investigated oblique shock-boundary layer interaction on a flat plate.
Understanding the physics of SBLI phenomena for the present configuration with wall curvature is essential to develop efficient supersonic fluid machinery, including the development of effective cooling strategies and active flow control for multi-point operation.

\section{Theoretical formulation and numerical methodology}
\label{section:numerical_methodology}

Wall-resolved LES are employed to solve the compressible Navier-Stokes equations written in terms of the contravariant velocity components.
Assuming a calorically perfect gas, the set of equations is closed by the equation of state.
The equations are solved in nondimensional form where the length, velocity components, density, pressure, temperature and time are nondimensionalized by the axial airfoil chord $c_{x}$, inlet speed of sound $a_{\infty}$, inlet density $\rho_{\infty}$, $\rho_{\infty} a_{\infty}^2$, $\left(\gamma - 1 \right) T_{\infty}$ and $c_{x}/a_{\infty}$, respectively. 
The Reynolds and Mach numbers are calculated as $Re = \rho_{\infty} U_{\infty} c_{x}/ \mu_{\infty}$ and $M_{\infty} = U_{\infty}/a_{\infty}$, respectively, where $U_{\infty}$, $T_{\infty}$ and $\mu_{\infty}$ represent the flow velocity, temperature and dynamic viscosity coefficient computed at the cascade inlet. 
The Prandtl number is given by $Pr = \mu_{\infty} c_{p} / \kappa_{\infty}$, where $c_p$ is the specific heat at constant pressure and $\kappa_{\infty}$ is the inlet thermal conductivity. The viscosity is computed using the nondimensional Sutherland law. 

The spatial discretization of the governing equations is performed using a sixth-order accurate compact scheme \citep{Nagarajan2003} implemented on a staggered grid. A sixth-order compact interpolation method is used to obtain fluid properties on the staggered nodes.
A sixth-order compact filter \cite{Lele1992} is applied in flow regions far away from solid boundaries at each time step to control numerical instabilities which may arise from mesh distortion and stretching, and interpolations between overlapping grids.

Two grids are employed in the present simulations: one is a body-fitted O-grid block which surrounds the blade and the other is a Cartesian block employed to facilitate the implementation of the pitchwise periodicity. In the O-grid, the time integration of the equations is carried out by the implicit second-order scheme of \citet{Beam1978} to reduce the stiffness problem typical of boundary layer grids. In the background Cartesian block, a third-order Runge-Kutta scheme is used for time advancement of the Navier-Stokes equations. A fourth-order Hermite interpolation scheme \citep{Bhaskaran} is used to exchange information between grid blocks in the overlapping zones. 
Further details about the numerical procedure can be found in Refs. \citep{Nagarajan2003, bhaskaran_thesis, Wolf2011}. The code has been previously validated for simulations of unsteady compressible flows \cite{bhaskaran_thesis,Wolf2012, Lui2022}, including the flow through turbine cascades \cite{Bhaskaran, Lui2022}. 

The localized artificial diffusivity LAD-D2-0 scheme  \citet{Kawai2010} is used to compute an artificial dissipation along shock waves. 
To promote the transition to turbulence on the airfoil boundary layers, an artificial body force is included in the momentum and energy equations, where a time-periodic unsteady actuation and random spanwise treatment are assumed. The forcing is applied along the wall-normal region up to a distance of $0.001c_{x}$ at $0.22 < x < 0.27 $ on the suction side, and at $0.10 < x < 0.15 $ on the pressure side, with the actuation changing every $\Delta t \approx$ 0.003 in a spanwise-random fashion.

\section{Flow and mesh configurations}
\label{section:configurations}

\begin{figure}
	\begin{overpic}[trim = 15mm 10mm 15mm 10mm, clip,width=0.99\textwidth]{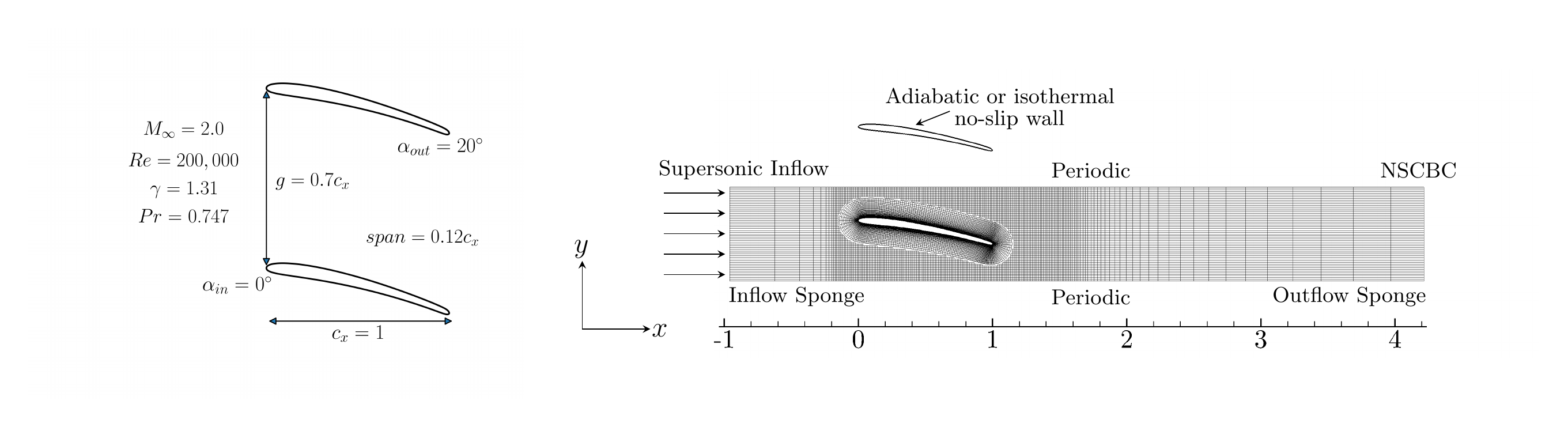}
		\put(0,18){(a)}
		\put(35,18){(b)}		
	\end{overpic} 
	\caption{Schematics of (a) flow configuration and geometrical parameters, and (b) computational domain skipping every $5$ grid points.}
	\label{fig:schematic}
\end{figure}

This section presents details of the flow configurations investigated and the computational grids employed in the LES calculations. Figure \ref{fig:schematic} (a) shows the flow conditions and geometrical parameters. The inlet Mach number is $M_{\infty}$ = 2.0 and the Reynolds number based on the inlet velocity and airfoil axial chord is $Re$ = 200,000. The fluid is assumed to be a calorically perfect gas, where the ratio of specific heats is $\gamma = 1.31$, the Prandlt number is $Pr = 0.747$ and the ratio of the Sutherland constant over inlet temperature is 0.07182. 
A realistic turbine cascade would be subjected to incoming turbulence. However, in the present analysis, the inlet consists of a ``clean'' inflow. This setup allows a more direct comparison against other SBLI studies available in the literature for canonical flow configurations.

Figure \ref{fig:schematic} (b) displays a schematic of the overset grid employed in the LES along with the implemented boundary conditions. The O-grid block has $1200 \times 280 \times 144$ points and is embedded in the background Cartesian grid block of size $960 \times 280 \times 72$. Therefore, the grid has approximately $68$ million points. Depending on the case, adiabatic or isothermal boundary conditions are applied along the blade surface. For the latter, the wall to inlet temperature ratio is $T_w/T_{\infty}=0.75$, representing a cooled wall. 
The boundary conditions are the supersonic inflow for the inlet and the Navier-Stokes characteristic boundary condition (NSCBC) \citep{Poinsot1992} for the outlet.
A damping sponge is also applied near the inflow and outflow boundaries to minimize reflections of numerical disturbances \citep{ISRAELI1981,Nagarajan2003}. 
Periodic boundary conditions are used in the $y$-direction of the background grid and in the spanwise direction, in order to simulate a linear cascade of blades and to enforce a statistically homogeneous flow along the span, respectively. The simulation is initialized with a uniform flow and statistics are computed after initial transients are discarded.
\begin{figure}[H]
\centering
\subfigure[Adiabatic]{\includegraphics[width=0.493\textwidth]{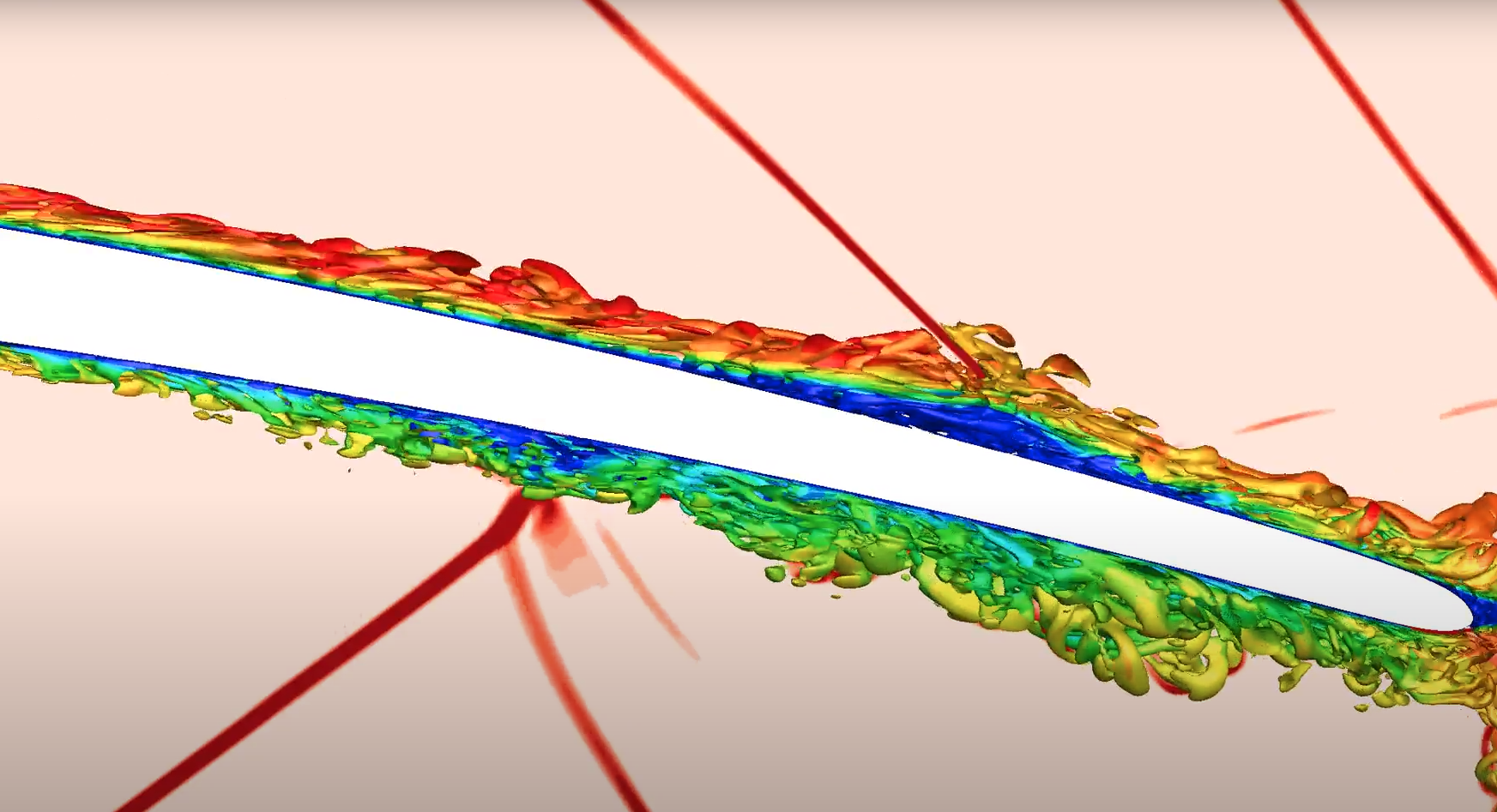}}
\subfigure[Cooled]{\includegraphics[width=0.493\textwidth]{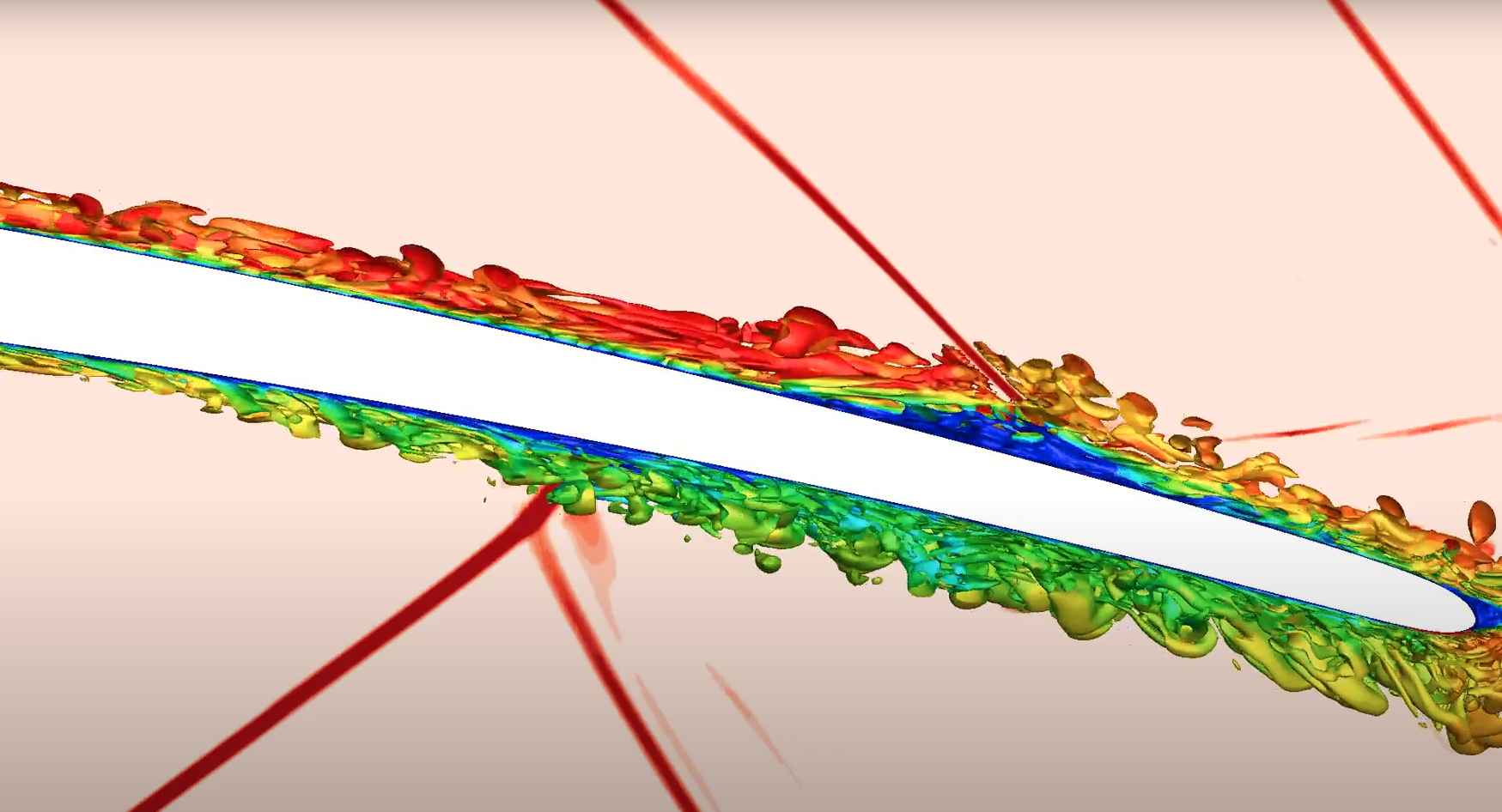}}
\caption{Iso-surfaces of $Q$-criterion colored by $u$-velocity component for different thermal boundary conditions. The background plane displays the shock waves by visualizing the density gradient magnitude $|\nabla \rho|$.}
\label{fig:flow_visualization}
\end{figure}

\section{Preliminary results}

Instantaneous flow snapshots are shown in Fig. \ref{fig:flow_visualization} with the iso-surfaces of $Q$-criterion colored by $u$-velocity to represent the turbulent structures along the airfoils. The magnitude of the density gradient is also shown in red to highlight the shock-waves. 
On the suction side, an oblique shock impinges on the turbulent boundary layer, causing the flow separation which is seen in blue.
A similar phenomenon is observed on the pressure side, but the incident shock becomes a Mach reflection that forms a normal shock near the wall. 
Figures \ref{fig:flow_visualization} (a) and (b) show detailed views of the separation bubbles for the adiabatic and cooled wall cases, respectively. It can be observed that cooling the wall results in a smaller recirculation regions on the suction side of the turbine.
\begin{figure}[H]
\centering

\subfigure[Adiabatic - $u$-velocity]{\includegraphics[trim={80mm 5mm 80mm 30mm}, clip, width=0.49\textwidth]{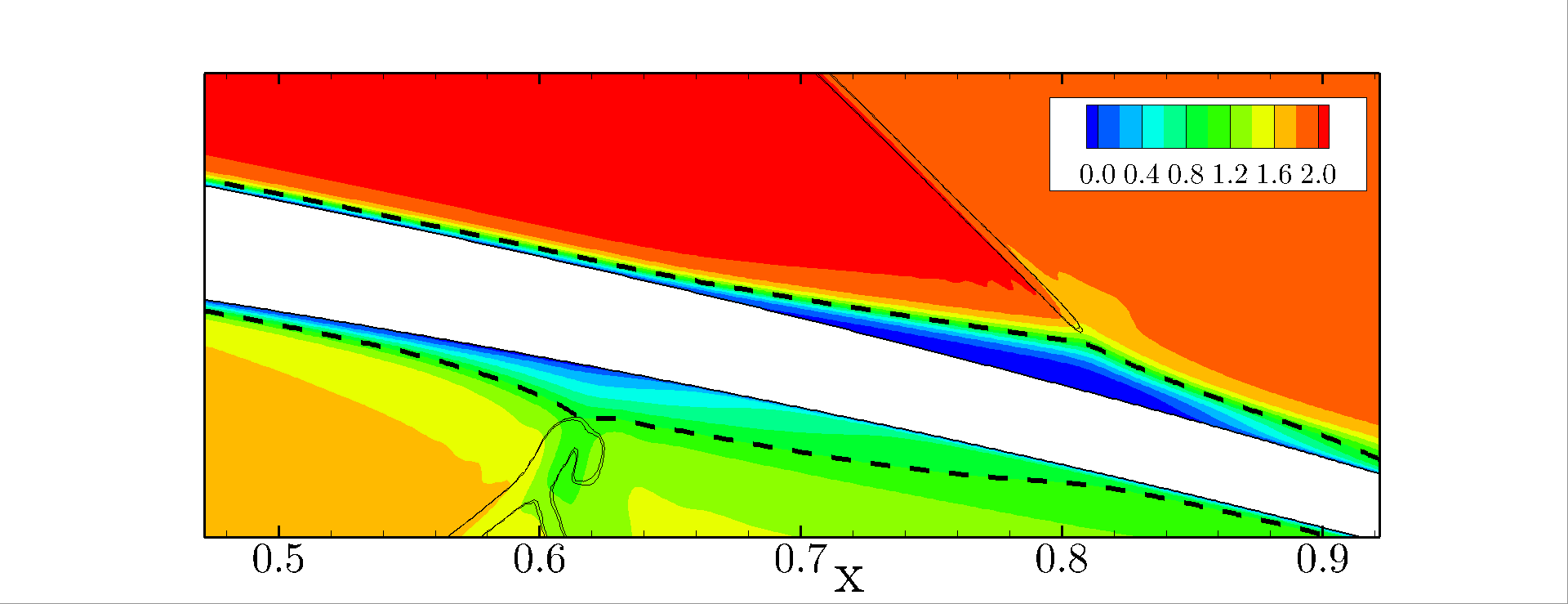}}
\subfigure[Cooled - $u$-velocity]{\includegraphics[trim={80mm 5mm 80mm 30mm}, clip, width=0.49\textwidth]{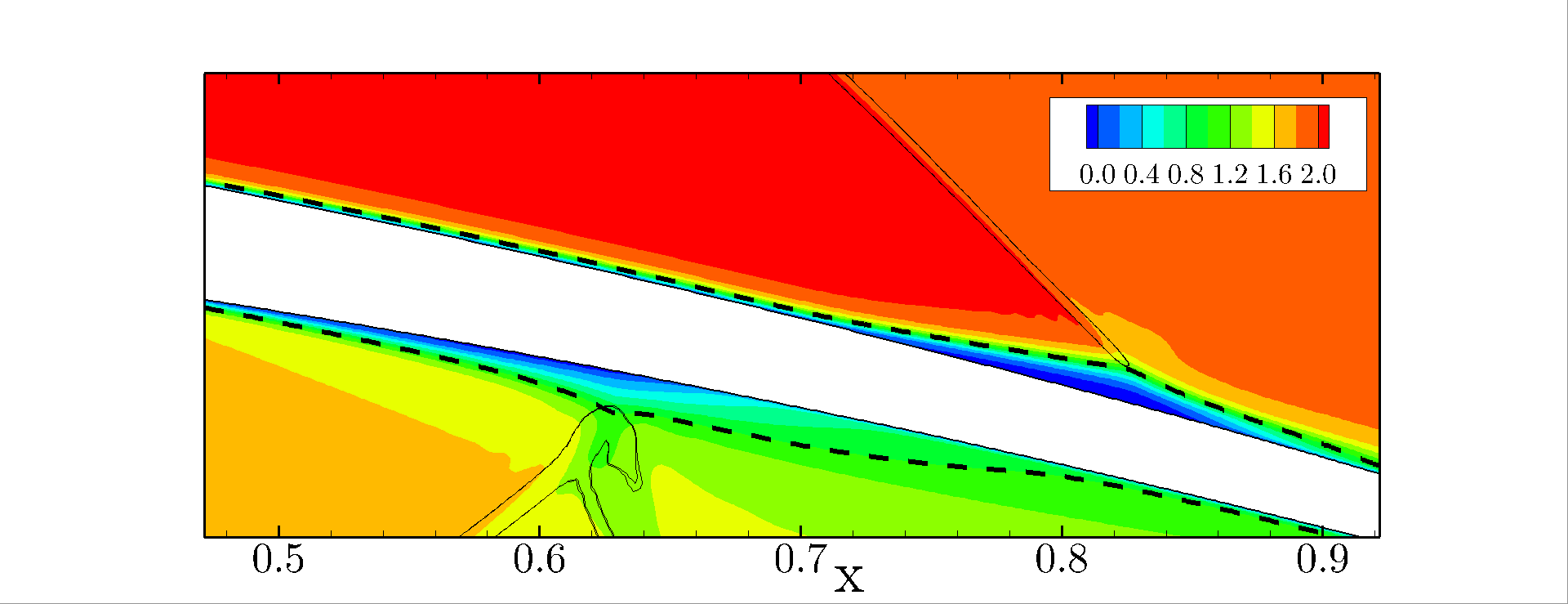}}

\subfigure[Adiabatic - suction side temperature]{\includegraphics[trim={80mm 5mm 80mm 30mm}, clip, width=0.49\textwidth]{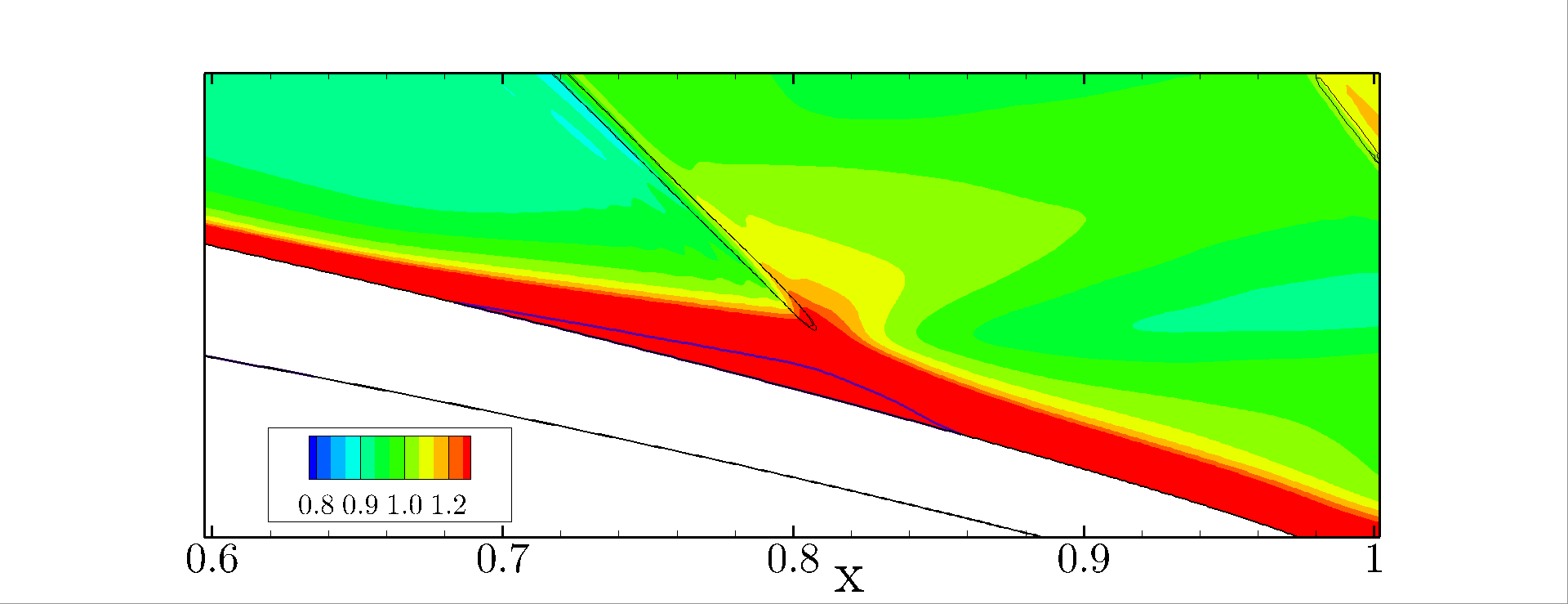}}
\subfigure[Cooled - suction side temperature]{\includegraphics[trim={80mm 5mm 80mm 30mm}, clip, width=0.49\textwidth]{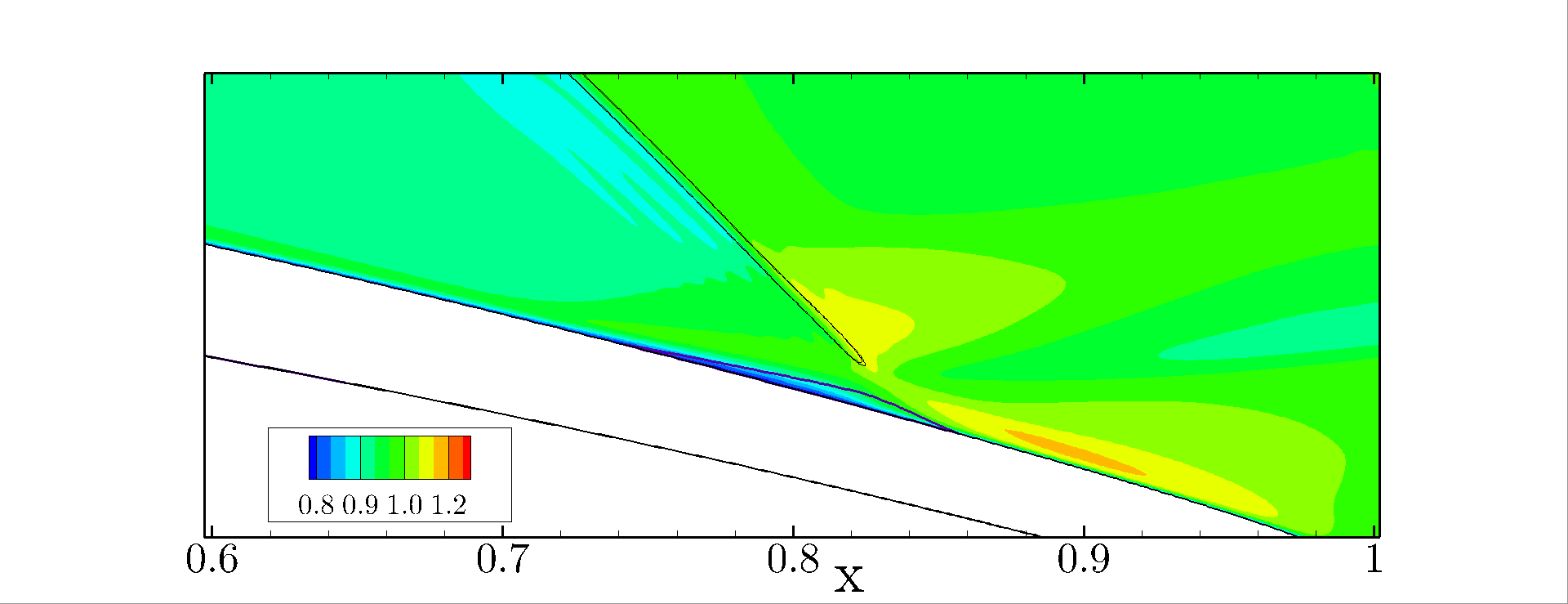}}

\subfigure[Adiabatic - pressure side temperature]{\includegraphics[trim={80mm 5mm 80mm 30mm}, clip, width=0.49\textwidth]{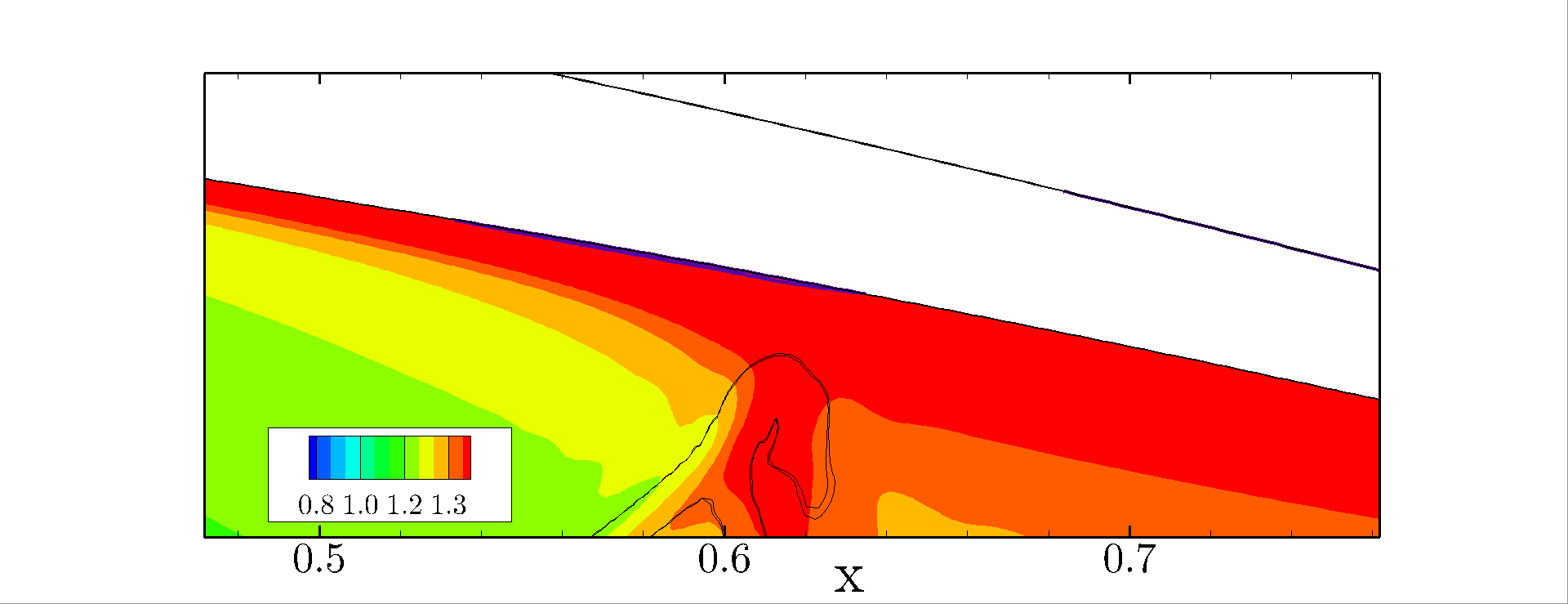}}
\subfigure[Cooled - pressure side temperature]{\includegraphics[trim={80mm 5mm 80mm 30mm}, clip, width=0.49\textwidth]{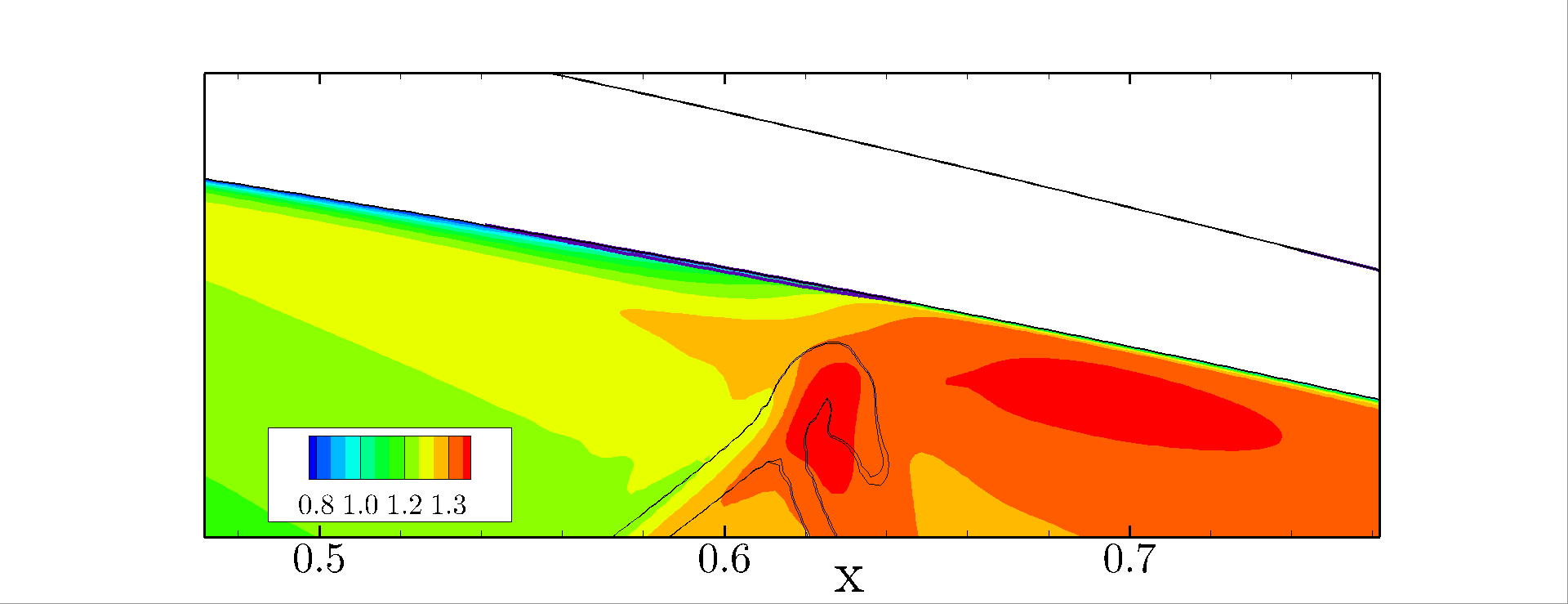}}

\caption{Spanwise and time-averaged contours of normalized $u$-velocity and temperature. The thicker black dashed lines show the sonic line and the thinner black lines display the shock waves visualized by the pressure gradient magnitude. Blue lines delimit the separation regions.}
\label{fig:mean_variables}
\end{figure}

To better visualize the separation bubbles, results are shown in terms of spanwise and time-averaged variables in Fig. \ref{fig:mean_variables}. Figures (a) and (b) show the contours of $u$-velocity for the adiabatic and cooled wall cases, respectively, allowing a comparison of the mean separation bubble length scales. It is possible to see that the cooled wall case has a smaller recirculation zone on the suction side of the turbine. Also, the sonic line is formed closer to the airfoil surface when cooling is applied. This results in a shock wave with a deeper penetration into the boundary layer, and this is also visible on the pressure side. 
Figures \ref{fig:mean_variables} (c) and (d) show the temperature contours for the suction side, while Figs. \ref{fig:mean_variables} (e) and (f) show the pressure side of the turbine. The adiabatic case  has its highest temperature near the surface and this is related to the absence of heat transfer to the wall, so the friction generated from the shear stresses are converted into heat and transported through the boundary layer. The cooled wall depicts the opposite behavior. The lowest values of temperature are observed at the wall and it is possible to see that the temperature rises abruptly downstream of the impinging shock waves, with the pressure side boundary layer displaying a more intense temperature rise. 

\begin{figure}
\centering
\subfigure[$\kappa$ - Both sides]{\includegraphics[trim={0cm 0cm 0cm 0cm},clip,width=0.32\textwidth]{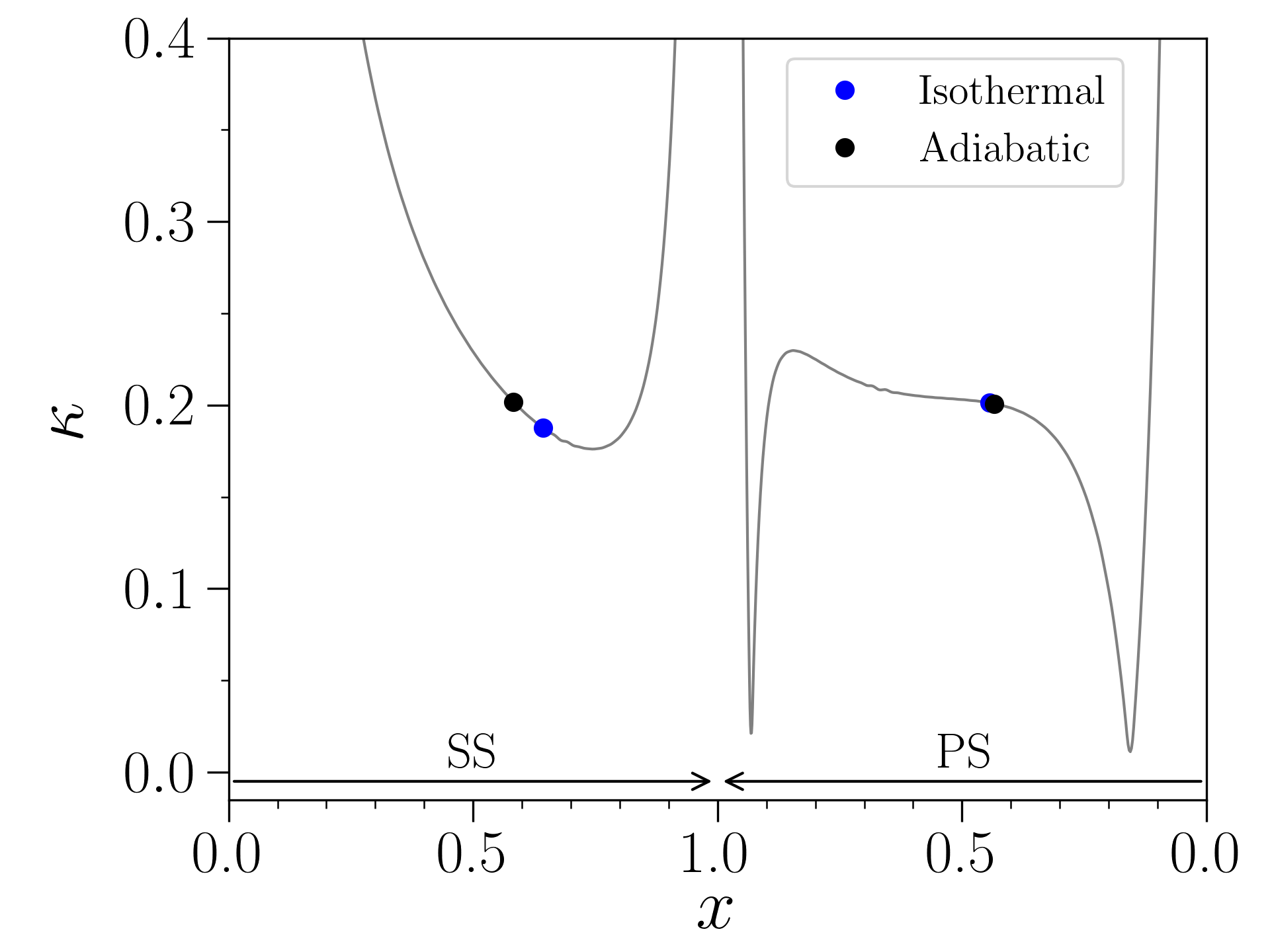}}
\subfigure[$\beta$ - Suction side]{\includegraphics[trim={0cm 0cm 0cm 0cm},clip,width=0.32\textwidth]{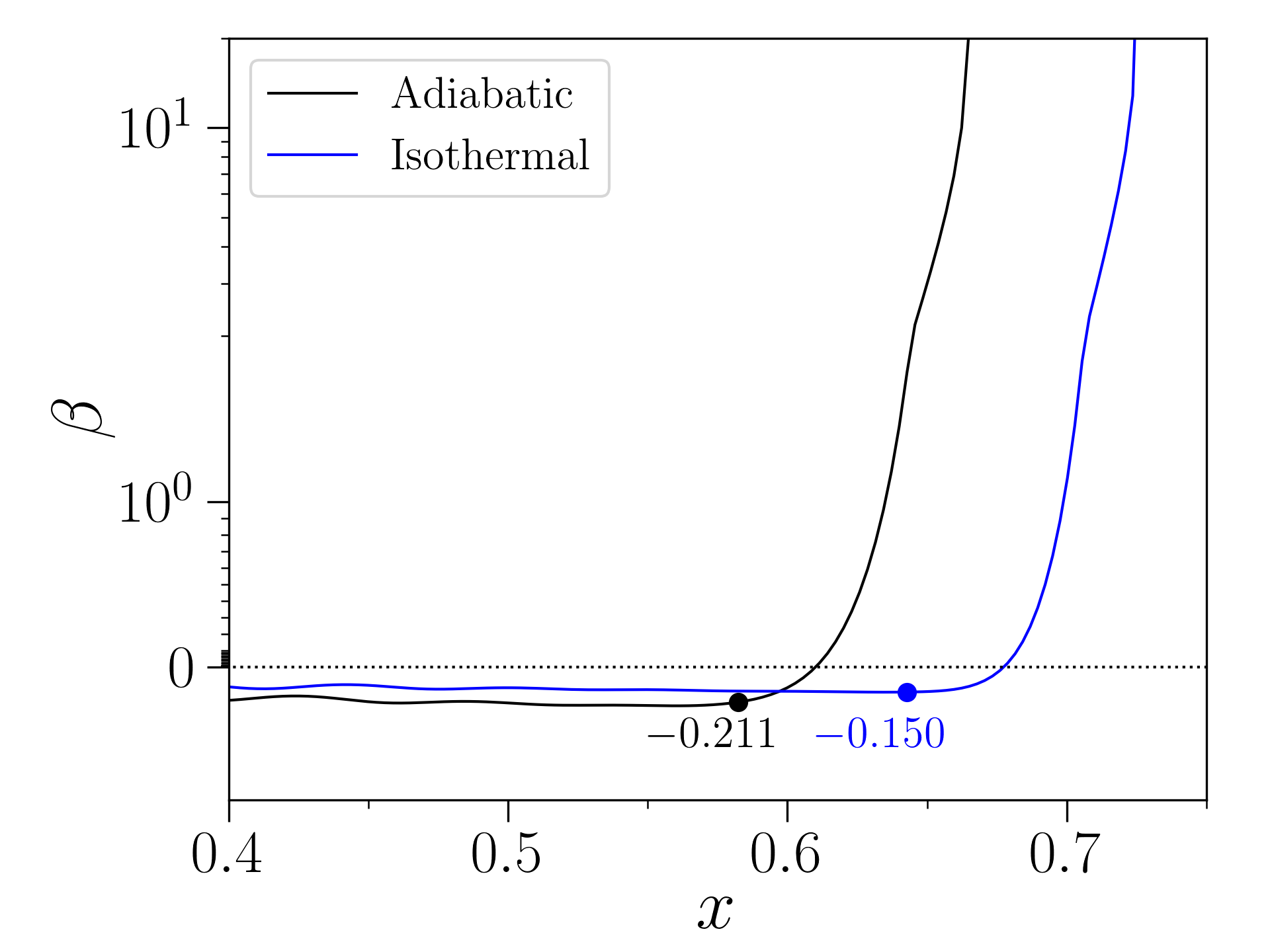}}
\subfigure[$\beta$ - Pressure side]{\includegraphics[trim={0cm 0cm 0cm 0cm},clip,width=0.32\textwidth]{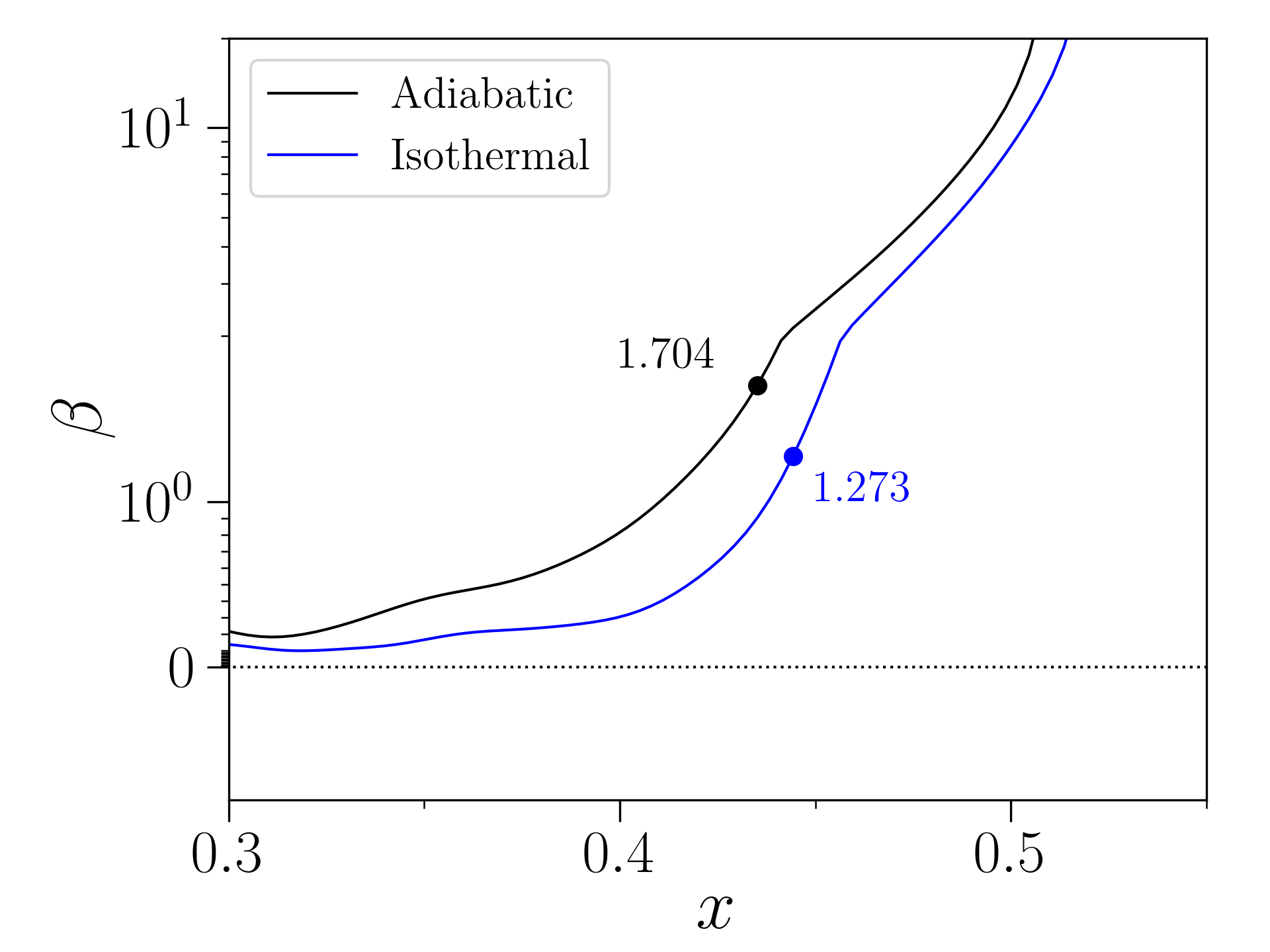}}

\caption{Chordwise distributions of the (a) airfoil curvature and the Clauser pressure gradient parameter on the (b) suction side and (c) pressure side. The marked points highlight positions located $10\%$ upstream of the bubble for each boundary condition.}
\label{fig:curvature_clauser}
\end{figure}

\begin{figure}
\centering
\subfigure[$\theta,\delta^*$ - Suction side]{\includegraphics[trim={0cm 0cm 0cm 0cm},clip,width=0.49\textwidth]{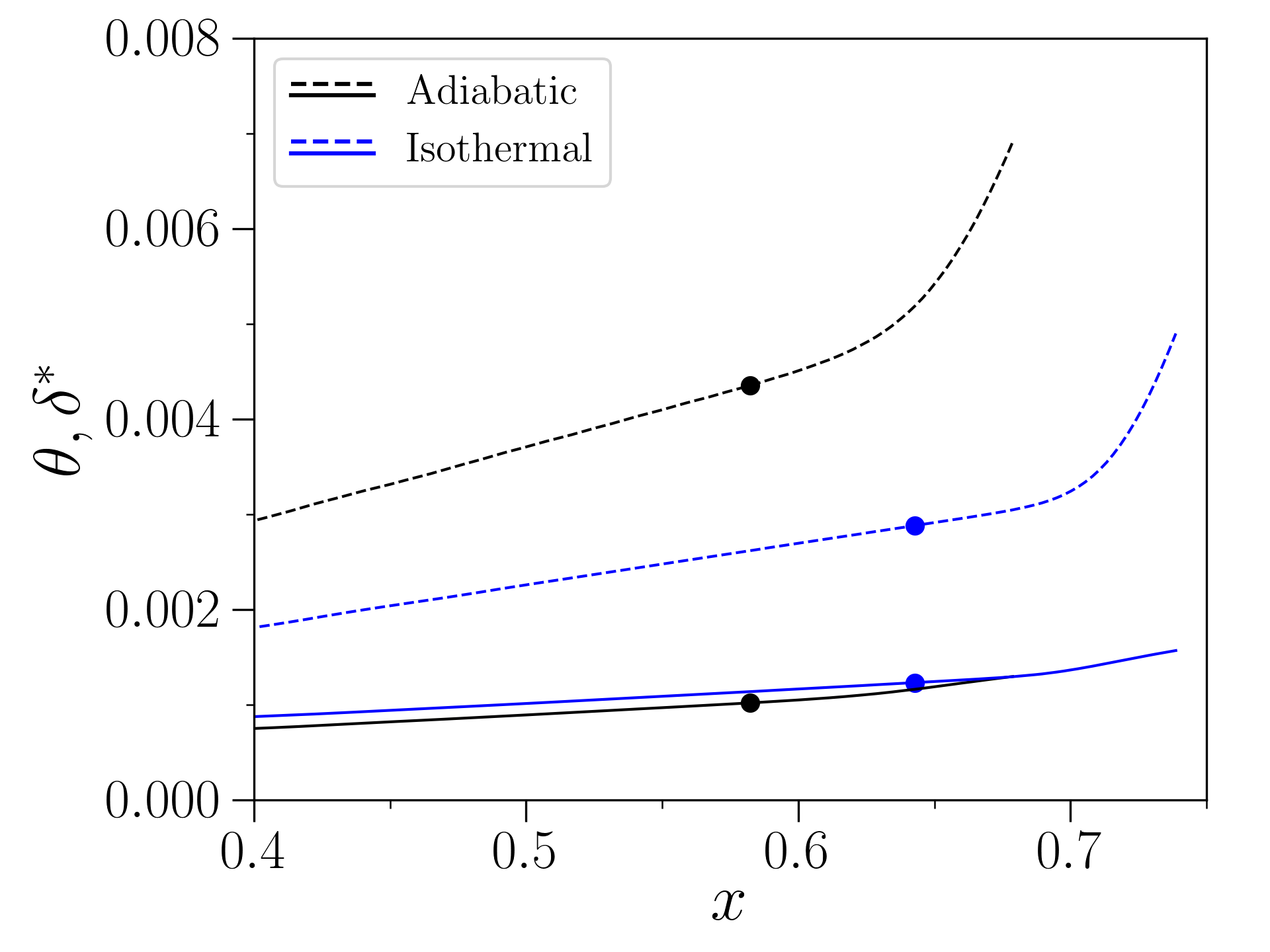}}
\subfigure[$\theta,\delta^*$ - Pressure side]{\includegraphics[trim={0cm 0cm 0cm 0cm},clip,width=0.49\textwidth]{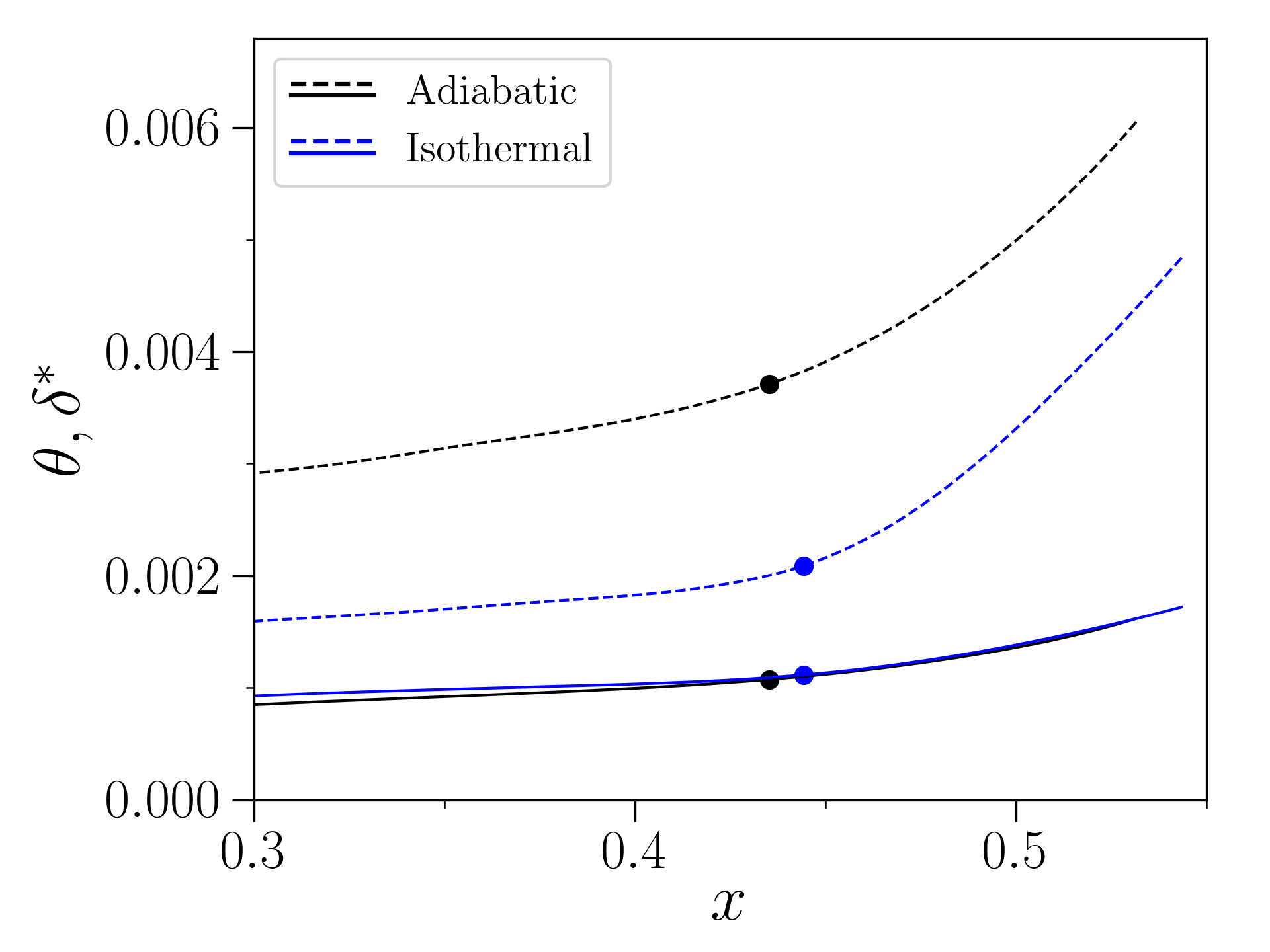}}

\subfigure[$H$ - Suction side]{\includegraphics[trim={0cm 0cm 0cm 0cm},clip,width=0.49\textwidth]{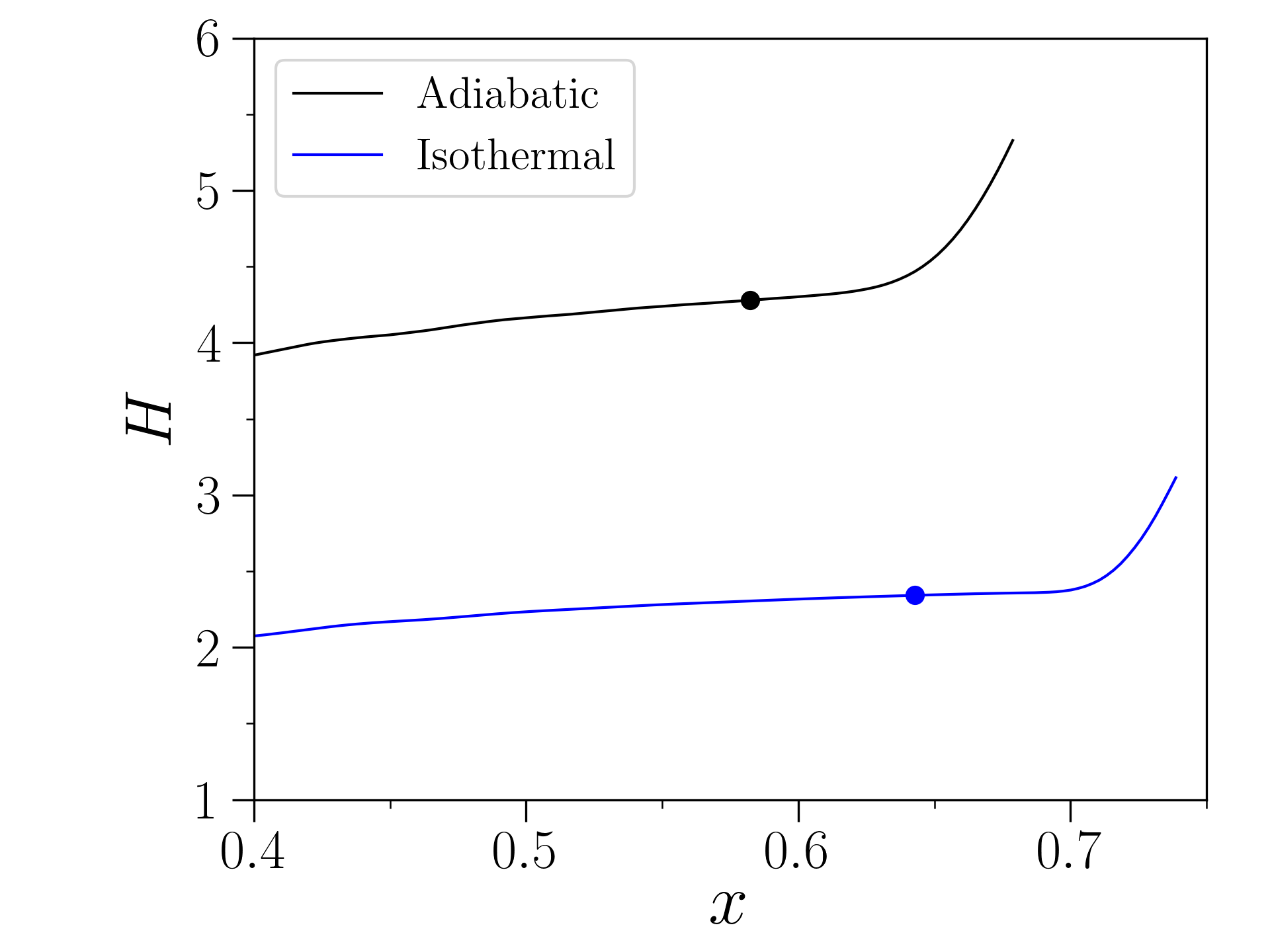}}
\subfigure[$H$ - Pressure side]{\includegraphics[trim={0cm 0cm 0cm 0cm},clip,width=0.49\textwidth]{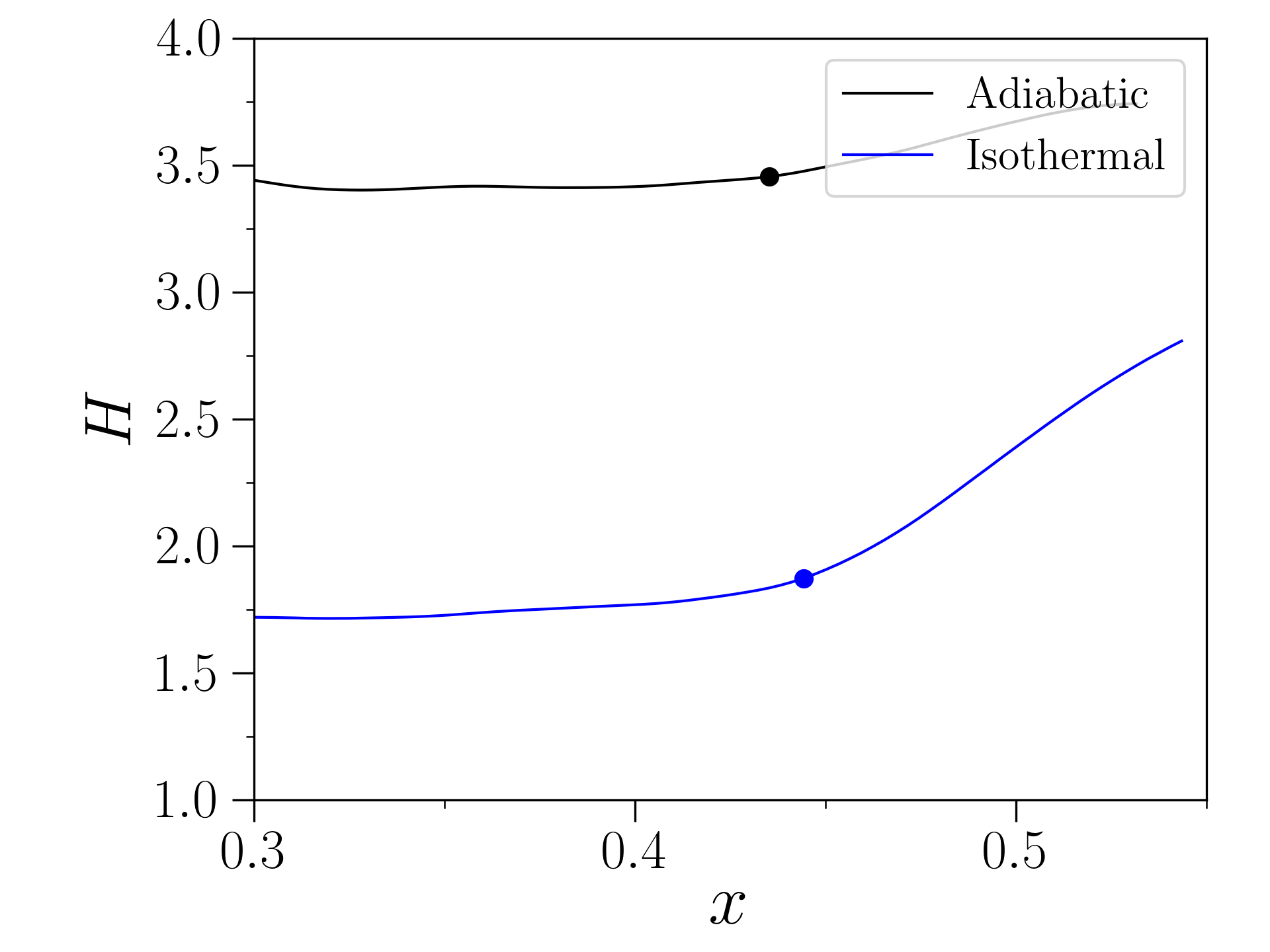}}
\caption{Chordwise distribution of boundary layer integral properties for the suction side (left) and pressure side (right). (a,b) displacement thickness $\delta^*$ (dashed lines) and momentum thickness $\theta$ (solid lines). (c,d) shape factor.}
\label{fig:thickness_shape}
\end{figure}

The incoming boundary layers on both sides of the airfoil are characterized in terms of pressure gradient and integral quantities in Figs. \ref{fig:curvature_clauser} and \ref{fig:thickness_shape}, respectively. The airfoil surface curvature $\kappa=1/R$ is shown in Fig. \ref{fig:curvature_clauser} (a) for reference, where $R$ is the local radius of curvature. 
The marked points in the figure highlight positions located 10\% upstream of the bubble
for each thermal boundary condition.
On the suction side, the curvature decreases in the flow direction until about $80\%$ of the chord, followed by a sharp increase on approaching the trailing edge. The flow develops further downstream under the influence of a lower curvature for the isothermal case, due to the delayed separation. For the pressure side, away from the leading and trailing edges, the curvature slightly increases in the flow direction (right to left in the plot). For this case, both boundary layers develop under a similar curvature, since the separation occurs nearly at the same point for the isothermal and adiabatic walls. 

The Clauser parameter \cite{clauser1954turbulent} provides information of the local pressure gradient including flow history effects, being defined as $\beta = (\delta^*/\tau_w)(dp_w/dx_t)$, where $\delta^*$ is the displacement thickness, $\tau_w$ is the wall shear stress, and $dp_w/dx_t$ is the wall pressure gradient in the flow direction. Chordwise distributions of $\beta$ until the flow separation are shown in Figs. \ref{fig:curvature_clauser} (b) and (c) for the suction and pressure sides, respectively. In the suction side, the boundary layer develops under a favorable pressure gradient ($\beta < 0$) until it experiences an abrupt rise, leading to an adverse pressure gradient, when reaching the compression waves upstream of the separation bubble. For the pressure side, the boundary layer develops under an adverse pressure gradient ($\beta >0$) and it experiences the same rise, but less abruptly, due to the compression waves. 
The effect of the adverse (favorable) pressure gradient in supersonic turbulent boundary layers is to increase (decrease) the wall shear stress and to decrease (increase) the boundary layer thickness \cite{spina1994physics}. 
So, in the suction side, the adiabatic boundary condition leads to a thicker boundary layer since the pressure gradient is more favorable than the isothermal case.
The opposite trend occurs on the pressure side, so the boundary layer should be thicker for the isothermal case if only the pressure gradient effects are taken into consideration.

The integral quantities are defined as: displacement thickness $\delta^* = \int^{\delta_e}_{0}[1-(\overline{\rho}/\rho_e)(\overline{u}_t/U_e)]dy_n$, momentum thickness $\theta = \int^{\delta_e}_{0}(\overline{\rho}/\rho_e)(\overline{u}_t/U_e)[1-(\overline{u}_t/U_e)]dy_n$, and shape factor $H=\delta^*/\theta$, where $u_t$ is the tangential velocity, the subscript $e$ denotes the flow quantities at the edge of the boundary layer, and the overline symbol corresponds to a mean quantity. The boundary layer thickness $\delta_e$ is defined as the wall normal location $y_n$ at which the mean spanwise vorticity $\omega_z$ falls below a suitable threshold value ($\omega_z \pm 10U_\infty/c_x$). Values of chordwise $\delta^*, \theta$ and $H$ are shown in Figs. \ref{fig:thickness_shape} (a,c) for the suction side and \ref{fig:thickness_shape} (b,d) for the pressure side. Regardless of the airfoil side, $\delta^*$ and $\theta$ grow in the flow direction, with an sharp rise for the former near the separation point. 
The suction and pressure sides of the airfoil are subjected to favorable and adverse pressure gradient conditions, as seen before. In this context, the analysis of $\beta$ alone is not sufficient to justify the differences in $\delta^*$ for both cases. The main difference between the two thermal boundary conditions is the temperature near the wall. For the adiabatic case, the higher wall temperature (Figs. \ref{fig:mean_variables} (c,e))  reduces the fluid density near the wall, leading to a mass flux towards the outer region \cite{spina1994physics} which thickens the boundary layer when compared to the isothermal case, which has a lower temperature near the wall (Figs. \ref{fig:mean_variables} (d,f)). This justifies the thicker displacement thickness for the adiabatic cases on both the suction and pressure sides. 

The level of mean kinetic energy of the velocity profile upstream of the SBLI tends to determine the response of the boundary layer when subjected to the adverse pressure gradient caused by the incident shock. A higher kinetic energy, represented by a fuller velocity profile, increases the resistance to separation of the boundary layer \cite{DELERY1985}. 
Based on this, the shape factor is used to quantify the fullness of the velocity profile, where a higher $H$ indicates a greater likelihood of flow separation. The chordwise distributions of the shape factor are shown in Figs. \ref{fig:thickness_shape} (c,d) for both sides of the airfoil. Regardless of the side, the adiabatic case has a higher $H$, indicating that it is more prone to separate. This is consistent with the results on the suction side, for which the separation occurs further upstream for the adiabatic case. For the pressure side, even though the adiabatic case is more prone to separate, the separation for both boundary layers occurs nearly at the same place, suggesting that another mechanism is responsible for the separation. 

\begin{figure}
\centering
\subfigure[]{\includegraphics[trim={0cm 0cm 0cm 0cm},clip,width=0.49\textwidth]{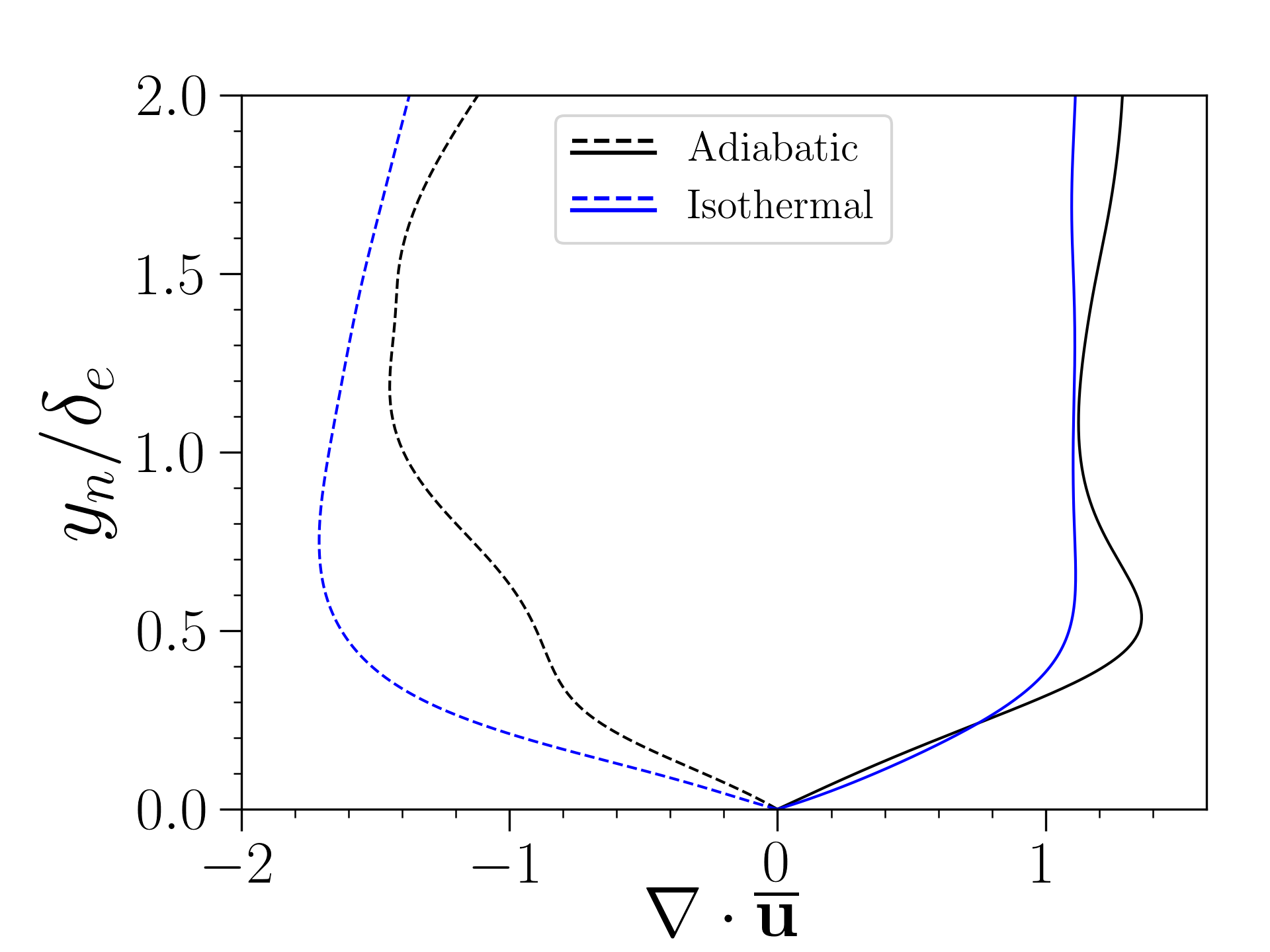}}
\subfigure[]{\includegraphics[trim={0cm 0cm 0cm 0cm},clip,width=0.49\textwidth]{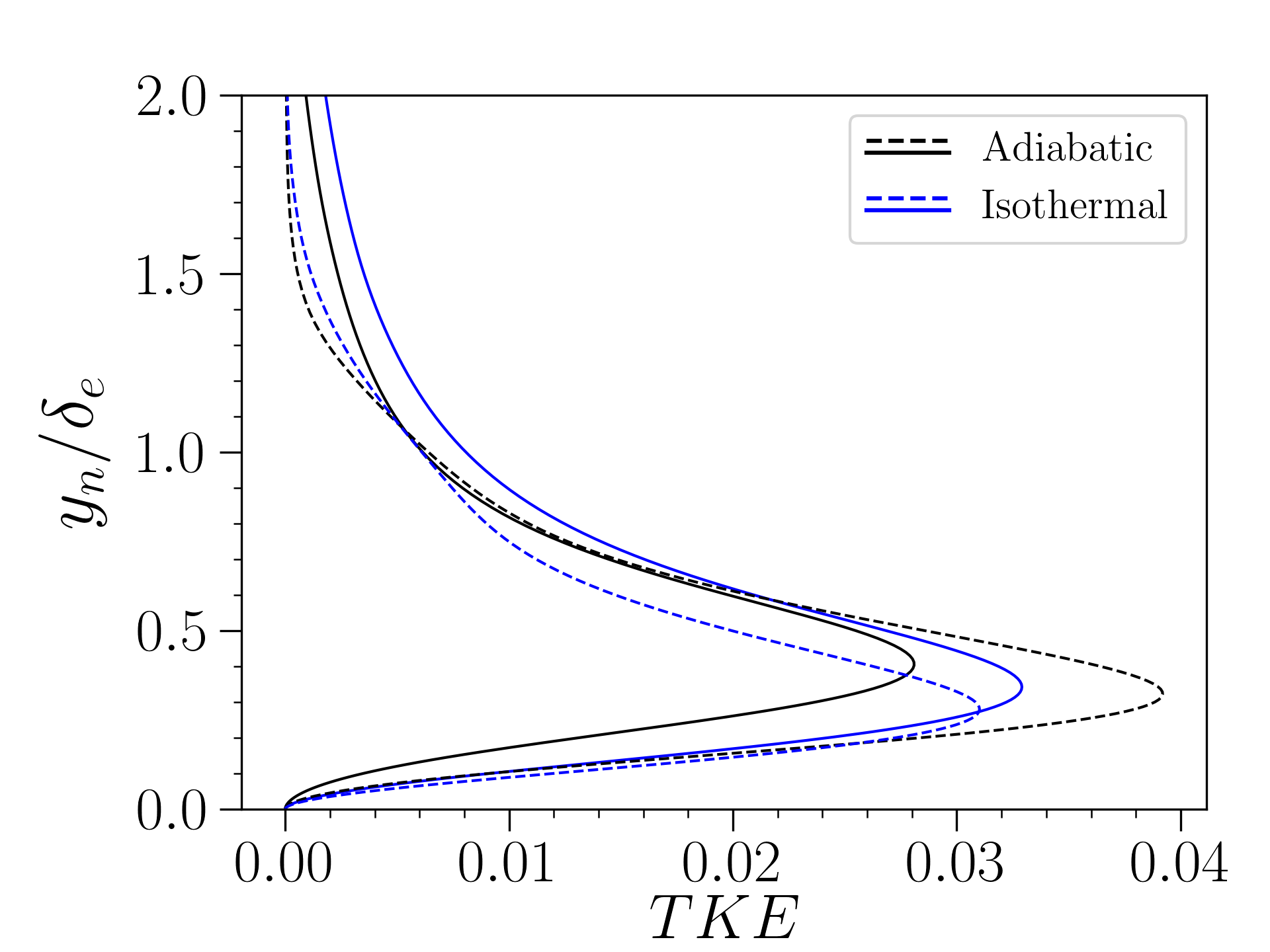}}

\caption{Profiles of (a) divergence of mean velocity and (b) turbulent kinetic energy at the reference positions ${\bf10\%}$ upstream of the separation bubble for each boundary condition. Solid (dashed) lines represent solutions for the suction (pressure) side.}
\label{fig:divVtke}
\end{figure}

The pressure side bubble is smaller than the suction side one based on the combined effects of pressure gradients, dilatation and TKE distribution. Profiles of the divergence of mean velocity ($\nabla\cdot\overline{u}$) and TKE are shown in Fig. \ref{fig:divVtke} at the reference positions located $10\%$ upstream of the separation bubbles.
On the pressure side, with its concave walls, the effects of the adverse pressure gradient and bulk compression (negative values of $\nabla\cdot\overline{u}$) imposed on the supersonic turbulent boundary layer results ultimately in an increased resistance to the boundary layer separation.
On the other hand, the supersonic boundary layer on the suction side is subjected to a favorable pressure gradient and bulk dilatation (positive values of $\nabla\cdot\overline{u}$) due to the convex wall. The aforementioned flow distortions introduce a stabilizing (destabilizing) extra strain rate in the suction (pressure) side boundary layer as discussed by Spina (1994) \cite{spina1994physics}. The destabilizing effect however results in turbulence amplification near the wall, which would increase the momentum transfer on the pressure side compared to the suction side. This can be seen in Fig. \ref{fig:divVtke} (b), which shows higher TKE values near the wall for the pressure side. The more intense turbulence activity near the wall leads to fuller velocity profiles which are less prone to separation and, thus, smaller bubbles on the pressure side.

When comparing the thermal boundary conditions, it is possible to see that the cooled case has a higher bulk compression on the pressure side (more unstable) and a lower bulk dilatation on the suction side (less stable). Moreover, the cooled case can be interpreted as having a higher local Reynolds number since the lower temperature near the wall results in a lower viscosity and, hence, higher turbulence levels. For the TKE profile, on the suction side, the cooled case has its peak closer to the wall and its intensity is higher, corroborating with the bulk dilatation and shape factor results. On the other hand, for the pressure side, even though the bulk compression and shape factor suggest a higher turbulence production for the cooled case, the opposite is observed. This phenomenon will be investigated in the final version of the paper. 

\section{Conclusion and additions to the final paper}

Wall-resolved LES are performed to investigate the effects of adiabatic and isothermal boundary conditions on SBLIs and the turbulent boundary layer characteristics for a supersonic turbine cascade. 
The simulations are performed for an inlet Mach number of $M_\infty = 2.0$ and Reynolds number $Re = 200,000$. For the isothermal case, the wall to inlet temperature ratio is $T_w/T_{\infty}=0.75$, representing a cooled surface. 
Flow snapshots are shown in terms of iso-surfaces of $Q$-criterion colored by $u$-velocity to visualize the separation regions, with the density gradient magnitude displaying the shock waves. 
The spanwise and time-averaged $u$-velocity contours are also presented, indicating that the shock wave for the cooled wall has a deeper penetration into the boundary layer due to the displacement of the sonic line. This leads to a steeper pressure variation of the incident shock. 

The temperature contours display a rise in temperature downstream of the shock impingement for the isothermal case, while for the adiabatic wall, the temperature near the surface is higher along the entire boundary layer due to the absence of heat conduction to the wall.
The airfoil surface curvature is shown for reference, with marked points highlighting the positions located $10\%$ upstream of the separation bubble for each boundary condition. 
On the suction side, the flow develops under the influence of a lower curvature for a longer distance for the isothermal case, since the flow separation occurs further downstream. For the pressure side, both boundary layers develop under a similar curvature, since the separation occurs nearly at the same point. The Clauser parameter shows that the boundary layers develop under a favorable (adverse) pressure gradient on the suction (pressure) side. We show that the pressure gradient alone is not capable of explaining the differences in the bubble sizes. 
For the adiabatic case, the higher temperature near the wall reduces the density that leads to a mass flux towards the outer region, leading to a thicker boundary layer when compared to its isothermal counterpart.

Integral quantities show a thicker displacement thickness for the adiabatic case, which results in a higher shape factor, corroborating that the adiabatic case is more prone to separation than the isothermal case. For the suction side this is indeed observed, but for the pressure side the adiabatic and isothermal cases have 
separation regions in the same location, suggesting that another mechanism is responsible for this . 
The smaller bubble on the pressure side is impacted by combined effects of pressure gradient, dilatation and TKE generation. Profiles of divergence of mean velocity and TKE show that, on the pressure side, the effects of the adverse pressure gradient and bulk compression result in an increased resistance to the boundary layer separation. The divergence of velocity results corroborate with the mean boundary layer parameters, and also show a higher local Reynolds number effect for the isothermal case. For the TKE results, the suction side results agree with the pressure gradient and the mean boundary layer paremeter analyses. But, for the pressure side, despite the bulk compression and shape factor suggesting a higher turbulence production for the cooled case, the opposite is observed. This phenomenon will be investigated in the final version of the paper. 

\section*{Acknowledgments}
The authors acknowledge Fundação de Amparo à Pesquisa do Estado de São Paulo, FAPESP, for supporting the present work under research grants No. 2013/08293-7, 2019/26196-5, 2021/06448-0 and 2022/00464-6. We also acknowledge the financial support from the Air Force Office of Scientific Research, AFOSR, for supporting the present work under grant FA9550-23-1-0615. This work was granted access to the HPC resources of IDRIS, TGCC, and CINES under the allocation A0152A12067 made by GENCI.

\bibliography{sample}

\end{document}